\documentclass[sigconf]{acmart}
\usepackage{booktabs} 
\usepackage{amsmath}
\usepackage{amsfonts}
\usepackage{comment}
\usepackage{footmisc}
\usepackage{mathrsfs}
\usepackage{amssymb}
\usepackage{graphicx}
\usepackage{subfigure}
\usepackage{multirow}
\usepackage{algorithm}
\usepackage{algorithmic}
\usepackage{multicol}  
\usepackage{tikz}
\usetikzlibrary{bayesnet}
\usepackage{enumitem}
\usepackage{array}
\usepackage{float}
\usepackage{hyperref}

\copyrightyear{2018} 
\acmYear{2018} 
\setcopyright{acmcopyright}
\acmConference[RecSys '18]{Twelfth ACM Conference on Recommender Systems}{October 2--7, 2018}{Vancouver, BC, Canada}
\acmBooktitle{Twelfth ACM Conference on Recommender Systems (RecSys '18), October 2--7, 2018, Vancouver, BC, Canada}
\acmPrice{15.00}
\acmDOI{10.1145/3240323.3240374}
\acmISBN{978-1-4503-5901-6/18/10}
\begin{document}
\title{Deep Reinforcement Learning for Page-wise Recommendations}

\author{Xiangyu Zhao}
\affiliation{
	\institution{Data Science and Engineering Lab\\
		Michigan State University}}
\email{zhaoxi35@msu.edu}

\author{Long Xia}
\affiliation{
	\institution{JD.com}}
\email{xialong@jd.com}

\author{Liang Zhang}
\affiliation{
	\institution{JD.com}
}
\email{zhangliang16@jd.com}
\author{Zhuoye Ding}
\affiliation{
	\institution{JD.com}}
\email{dingzhuoye@jd.com}
\author{Dawei Yin}
\affiliation{
	\institution{JD.com}}
\email{yindawei@acm.org}
\author{Jiliang Tang}
\affiliation{
	\institution{Data Science and Engineering Lab\\
		Michigan State University}}
\email{tangjili@msu.edu}
\renewcommand{\shortauthors}{Xiangyu Zhao et al.}

\begin{abstract}
Recommender systems can mitigate the information overload problem by suggesting users' personalized items. In real-world recommendations such as e-commerce, a typical interaction between the system and its users is -- users are recommended a page of items and provide feedback; and then the system recommends a new page of items. To effectively capture such interaction for recommendations, we need to solve two key problems -- (1) how to update recommending strategy according to user's \textit{real-time feedback}, and 2) how to generate a page of items with proper display, which pose tremendous challenges to traditional recommender systems. In this paper, we study the problem of page-wise recommendations aiming to address aforementioned two challenges simultaneously. In particular, we propose a principled approach to jointly generate a set of complementary items and the corresponding strategy to display them in a 2-D page; and propose a novel page-wise recommendation framework based on deep reinforcement learning, DeepPage, which can optimize a page of items with proper display based on real-time feedback from users. The experimental results based on a real-world e-commerce dataset demonstrate the effectiveness of the proposed framework.
\end{abstract}
\maketitle

\section{Introduction} \label{sec:introduction}
\begin{figure}[t]
	\centering
	\includegraphics[width=81mm]{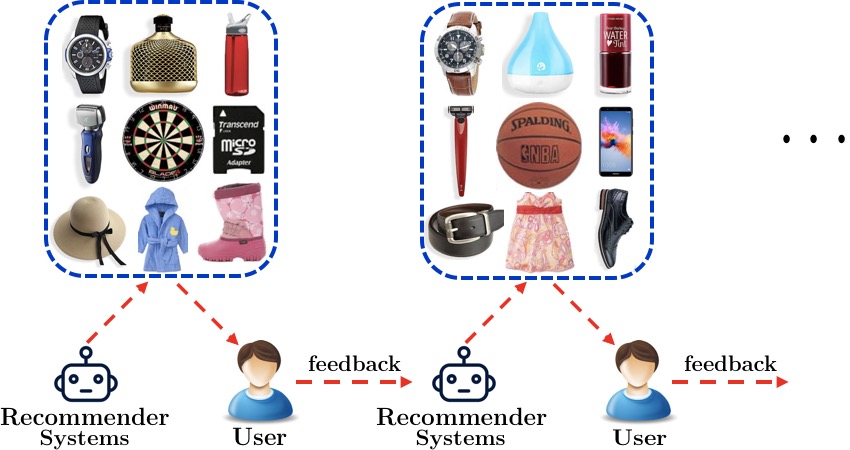}
	\caption{An example to illustrate the interactions between recommender systems and users.}
	\label{fig:interaction}
	\vspace{-4.9mm}
\end{figure}

Recommender systems are intelligent E-commerce applications~\cite{linden2003amazon,breese1998empirical,mooney2000content}. They assist users in their information-seeking tasks by suggesting items (products, services, or information) that best fit their needs and preferences. Recommender systems have become increasingly popular in recent years, and have been utilized in a variety of domains including movies, music, locations, and social events~\cite{resnick1997recommender,ricci2011introduction,zhao2016exploring,wang2017your,guo2016cosolorec,gao2015content,Bao2015Recommendations,Zheng2011Recommending}. Figure \ref{fig:interaction} illustrates a typical example of the interactions between an e-commerce recommender system and a user -- each time the system recommends a page of items to the user; next the user browses these items and provides real-time feedback and then the system recommends a new page of items. This example suggests two key challenges to effectively take advantage of these interactions for e-commerce recommender systems -- 1) how to efficiently capture user's preference and update recommending strategy according to user's real-time feedback; and 2) how to generate a page of items with proper display based on user's preferences.

\vspace{-3mm}
\subsection{Real-time Feedback}\label{sec:real-time}
Most existing recommender systems consider the recommendation procedure as a static process and make recommendations following a fixed greedy strategy. However, these approaches may fail in capturing the dynamic nature of the users' preferences, and they become infeasible to efficiently and continuously update their recommending strategies according to user's real-time feedback. Thus, in this work, we consider the recommendation procedure as sequential interactions between users and the recommender agent; and leverage Reinforcement Learning (RL) to automatically learn the optimal recommendation strategies. Recommender systems based on reinforcement learning have two major advantages. First, they are able to continuously update their strategies based on user's real-time feedback during the interactions, until the system converges to the optimal strategy that generates recommendations best fitting users' dynamic preferences. Second, the optimal strategy is made by maximizing the expected long-term cumulative reward from users; while the majority of traditional recommender systems are designed to maximize the immediate (short-term) reward of recommendations~\cite{shani2005mdp}. Therefore, the system can identify items with small immediate rewards but making big contributions to the rewards for future recommendations.
\vspace{-5mm}
\subsection{Page-wise Recommendations}
As mentioned in the example, users are typically recommended a page of items.  To achieve this goal, we introduce a page-wise recommender system, which is able to jointly (1) generate a set of diverse and complementary items and (2) form an item display strategy to arrange the items in a 2-D page that can lead to maximal reward. Conventional RL methods could recommend a set of items each time, for instance, DQN can recommend a set of items with highest Q-values according to the current state\cite{mnih2013playing}. However, these approaches recommend items based on the same state, which leads to the recommended items to be similar. In practice, a bundling of complementary items may receive higher rewards than recommending all similar items. For instance, in real-time news feed recommendations, a user may want to read diverse topics of interest\cite{yue2011linear}. In addition, page-wise recommendations need to properly display a set of generated items in a 2-D page. Traditional approaches treat it as a ranking problem, i.e., ranking items into a 1-D list according to the importance of items. In other words, user's most preferred item is posited in the top of list. However, in e-commerce recommender systems, a recommendation page is a 2-D grid rather than a 1-D list. Also eye-tracking studies~\cite{srikant2010user} show that rather than scanning a page in a linear fashion, users do page chunking, i.e., they partition the 2-D page into chunks, and browse the chunk they prefer more. In addition, the set of items and the display strategy are generated separately; hence they may be not optimal to each other. Therefore, page-wise recommendations need principled approaches to simultaneously generate a set of complementary items and the display strategy in a 2-D page. 
\vspace{-5mm}
\subsection{Contributions}
In this paper, we tackle the two aforementioned challenges simultaneously by introducing a novel page-wise recommender system based on deep reinforcement learning.  We summarize our major contributions as follows -- (1) we introduce a principled approach to generate a set of complementary items and properly display them in one 2-D recommendation page simultaneously;  (2) we propose a page-wise recommendation framework DeepPage, which can jointly optimize a page of items by incorporating real-time feedback from users; and (3) we demonstrate the effectiveness of the proposed framework in a real-world e-commerce dataset and validate the effectiveness of the components in DeepPage for accurate recommendations.


\vspace{-3mm}
\section{The Proposed Framework}\label{sec:framework}
In this section, we first give an overview of the proposed Actor-Critic based reinforcement learning recommendation framework with notations. Then we present the technical details of components in Actor and Critic, respectively.

\vspace{-3mm}
\subsection{Framework Overview}
\label{sec:problem} 
As mentioned in Section \ref{sec:real-time}, we model the recommendation task as a Markov Decision Process (MDP) and leverage Reinforcement Learning (RL) to automatically learn the optimal recommendation strategies, which can continuously update recommendation strategies during the interactions and the optimal strategy is made by maximizing the expected long-term cumulative reward from users. With the above intuitions, we formally define the tuple of five elements $(\mathcal{S}, \mathcal{A}, \mathcal{P}, \mathcal{R}, \gamma)$ of MDP --  (a) {\bf State space $\mathcal{S}$}: A state $s  \in \mathcal{S}$ is defined as user's current preference, which is generated based on user's browsing history, i.e., the items that a user browsed and her corresponding feedback; (b) {\bf Action space $\mathcal{A}$}:  An action $a = \{a^1,  \cdots, a^M\}\in \mathcal{A}$ is to recommend a page of $M$ items to a user based on current state $s$; (c) {\bf Reward $\mathcal{R}$}: After the RA takes an action $a$ at the state $s$, i.e., recommending a page of items to a user, the user browses these items and provides her feedback. She can skip (not click), click, or purchase these items, and the agent receives immediate reward $r(s,a)$ according to the user's feedback; (d) {\bf Transition $\mathcal{P}$}: Transition $p(s'|s,a)$ defines the state transition from $s$ to $s'$ when RA takes action $a$; and (e) {\bf Discount factor $\gamma$}: $\gamma \in [0,1]$ defines the discount factor when we measure the present value of future reward. In particular, when $\gamma = 0$, RA only considers the immediate reward. In other words, when $\gamma = 1$, all future rewards can be counted fully into that of the current action.

Specifically, we model the recommendation task as a MDP in which a recommender agent (RA) interacts with environment $\mathcal{E}$ (or users) over a sequence of time steps. At each time step, the RA takes an action $a \in \mathcal{A}$ according to $\mathcal{E}$'s state $s  \in \mathcal{S}$, and receives a reward $r(s,a)$ (i.e. the RA recommends a page of items according to user's current preference, and receives user's feedback). As the consequence of action $a$, the environment $\mathcal{E}$ updates its state to $s'$ with transition $p(s'|s,a)$. The goal of reinforcement learning is to find a recommendation policy $\pi:\mathcal{S} \rightarrow \mathcal{A}$, which can maximize the cumulative reward for the recommender system.

\begin{figure}[t]
	\centering
	\includegraphics[width=90mm]{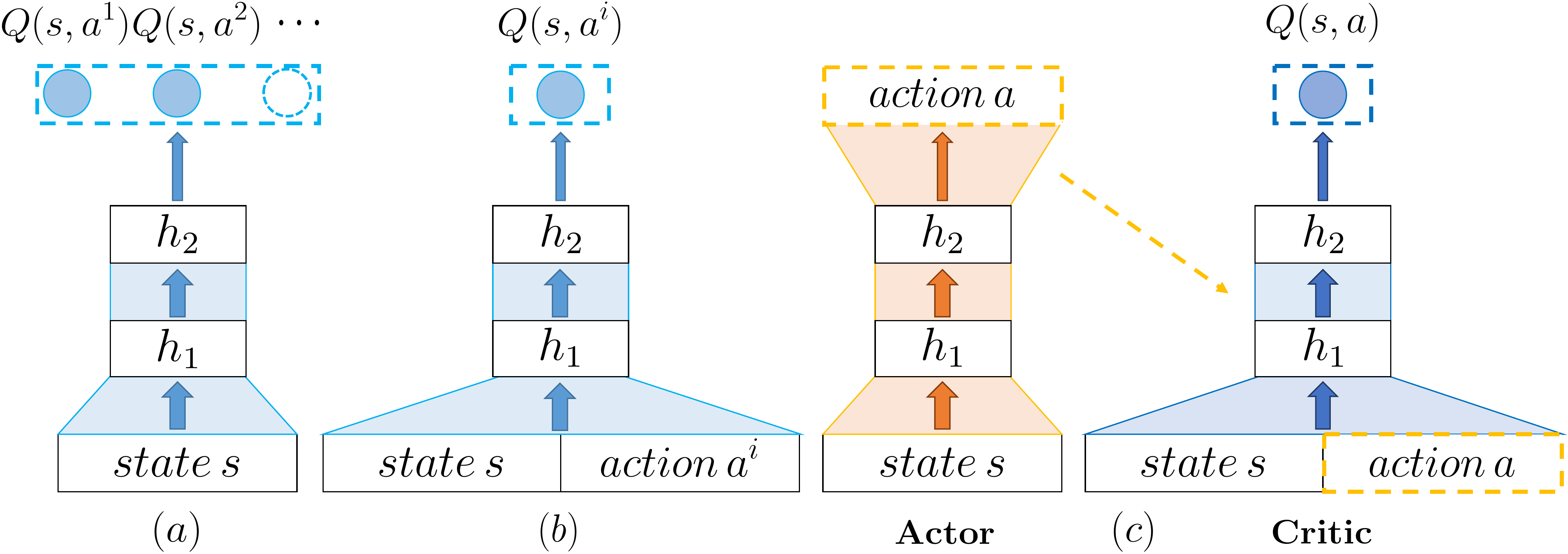}
	\caption{Framework architecture selection.}
	\label{fig:selection}
	\vspace{-10mm}
\end{figure}

In practice, conventional RL methods like Q-learning\cite{taghipour2007usage,taghipour2008hybrid} and POMDP\cite{shani2005mdp,hauskrecht1997incremental,ng2000pegasus,poupart2005vdcbpi,kearns2002sparse} become infeasible with the increasing number of items for recommendations. Thus, we leverage Deep Reinforcement Learning\cite{lillicrap2015continuous} with (adapted) artificial neural networks as the non-linear approximators to estimate the action-value function in RL. This model-free reinforcement learning method does not estimate the transition probability and not store the Q-value table. Hence it can support huge amount of items in recommender systems. 

There are two major challenges when we apply deep reinforcement learning to the studied problem -- (a) the large (or even continuous) and dynamic action space (item space), and (b) the computational cost to select an optimal action (a page of items). In practice, only using discrete indices to denote items is not sufficient since we cannot know the relations between different items only from indices. One common way is to use extra information to represent items with continuous embeddings\cite{levy2014neural}. Besides, the action space of recommender systems is dynamic as items are arriving and leaving. Moreover, computing Q-value for all state-action pairs is time-consuming because of the enormous state and action spaces. 

To tackle these challenges, in this paper, our recommending policy builds upon the Actor-Critic framework~\cite{sutton1998reinforcement}, shown in Figure~\ref{fig:selection} (c). The Actor-Critic architecture is preferred from the studied problem since it is suitable for large and dynamic action space, and can also reduce redundant computation simultaneously compared to alternative architectures as shown in Figures~\ref{fig:selection} (a) and (b).  The conventional Deep Q-learning architectures shown in Figure~\ref{fig:selection} (a) inputs only the state space and outputs Q-values of all actions. This architecture is suitable for the scenario with high state space and small/fixed action space like Atari\cite{mnih2013playing}, but cannot handle large and dynamic action space scenario, like recommender systems. Also, we cannot leverage the second conventional deep Q-learning architecture as shown in Fig.\ref{fig:selection}(b) because of its temporal complexity. This architecture inputs a state-action pair and outputs the Q-value correspondingly, and makes use of the optimal action-value function $Q^*(s, a)$. It is the maximum expected return achievable by the optimal policy, and should follow the Bellman equation \cite{bellman2013dynamic} as:
\begin{small}
\begin{equation}\label{equ:Q*sa}
\vspace{-1mm}
Q^{*}(s, a)=\mathbb{E}_{s'} \, \big[r+\gamma\max_{a'}Q^{*}(s', a')|s, a\big].
\end{equation}
\end{small}
In practice, selecting an optimal $a'$, $|\mathcal{A}|$ evaluations is necessary for the inner operation ``$\max_{a'}$''. In other words, this architecture computes Q-value for all $a' \in \mathcal{A}$ separately, and then selects the maximal one. This prevents Equation (\ref{equ:Q*sa}) from being adopted in practical recommender systems. 

In the Actor-Critic framework, the Actor architecture inputs the current state $s$ and aims to output a deterministic action (or recommending a deterministic page of $M$ items), i.e., $s \rightarrow a = \{a^1,  \cdots, a^M\}$. The Critic inputs only this state-action pair rather than all potential state-action pairs, which avoids the aforementioned computational cost as follows:
\begin{small}
\begin{equation}\label{equ:Qsa}
\vspace{-1mm}
Q(s, a)=\mathbb{E}_{s'} \, \big[r+\gamma Q(s', a')|s, a\big],
\end{equation}
\end{small}
where the Q-value function $Q(s, a)$ is a judgment of whether the selected action matches the current state, i.e., whether the recommendations match user's preference. Finally, according to the judgment from Critic, the Actor updates its' parameters in a direction of boosting recommendation performance so as to output properer actions in the next iteration.  Next we will elaborate the Actor and Critic architectures.

\vspace{-2mm}
\subsection{Architecture of Actor Framework}
\label{sec:actor}
The Actor is designed to generate a page of recommendations according to user's preference, which needs to tackle three challenges -- 1) setting an initial preference at the beginning of a new recommendation session, 2) learning the real-time preference in the current session, which should capture the dynamic nature of user's preference in current session and user's preferable item display patterns in a page, and 3) jointly generating a set of recommendations and displaying them in a 2-D page. To address these challenges, we propose an Actor framework with the Encoder-Decoder architecture. 

\begin{figure}[t]
	\centering
	\includegraphics[width=81mm]{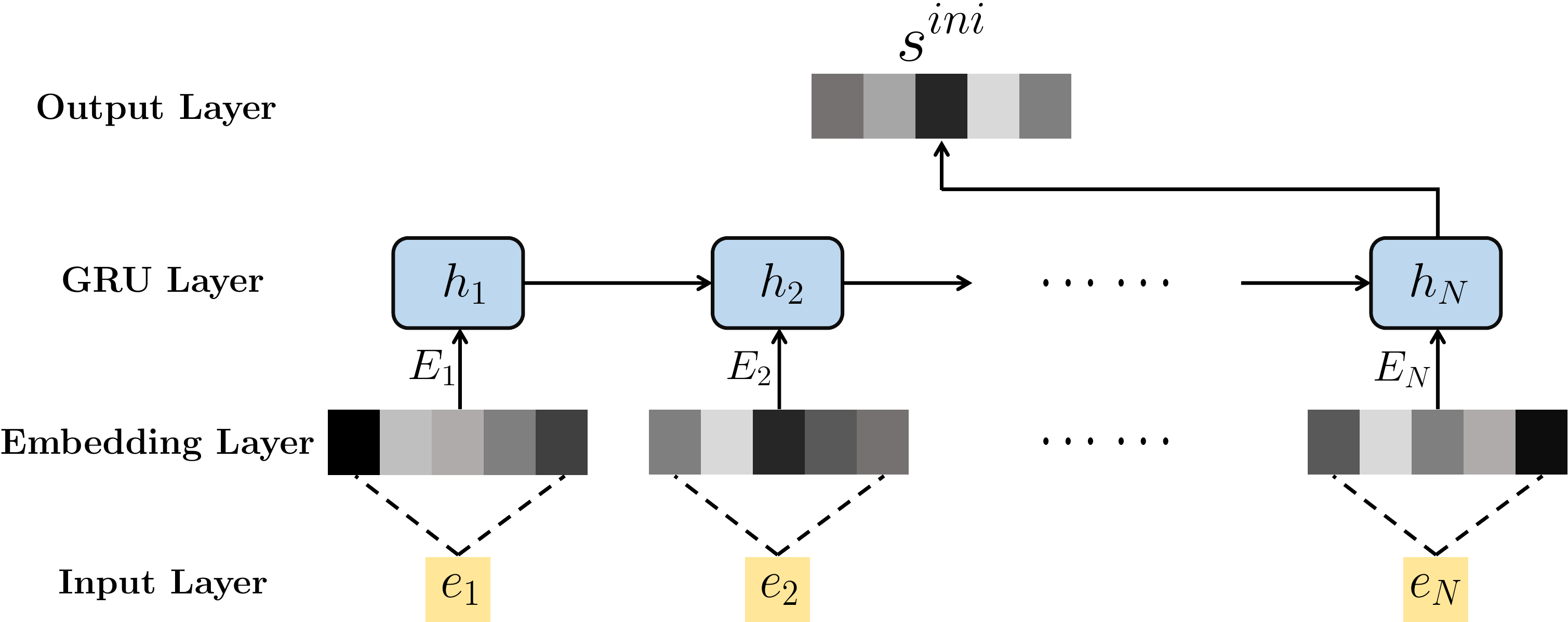}
	\caption{Encoder to generate initial state $s^{ini}$.\label{fig:gru1}}
	\vspace{-6mm}
\end{figure}

\vspace{-2mm}
\subsubsection{\textbf{Encoder for Initial State Generation Process}} 
\label{sec:initial}
Figure~\ref{fig:gru1} illustrates the model for generating initial preference. We introduce a RNN with Gated Recurrent Units (GRU) to capture users' sequential behaviors as user's initial preference. The inputs of GRU are user's last clicked/purchased items $\{e_1, \cdots, e_{N}\}$ (sorted in chronological order) before the current session, while the output is the representation of users' initial preference by a vector. The input $\{e_1, \cdots, e_{N}\}$ is dense and low-dimensional vector representations of items~\footnote{These item representations are pre-trained using users' browsing history by a company, i.e.  each item is treated as a word and the clicked items in one recommendation session as a sentence, and item representations are trained via word embedding\cite{levy2014neural}. The effectiveness of these item representations is validated by their business such as searching, ranking, bidding and recommendations. }. We add an item-embedding layer to transform $e_i$ into a low-dimensional dense vector via $E_i = \tanh(W_E e_i + b_E)\in \mathbb{R}^{|E|}$ where we use ``$\tanh$'' activate function since $e_i \in (-1,+1)$.

We leverage GRU rather than Long Short-Term Memory (LSTM) because that GRU outperforms LSTM for capturing users' sequential preference in some recommendation tasks \cite{hidasi2015session}. Unlike LSTM using input gate $i_t$ and forget gate $f_t$ to generate a new state, GRU utilizes an update gate $z_t$:
\begin{small}
\begin{equation}\label{equ:update}
z_{t}=\sigma(W_{z}E_{t}+U_{z}h_{t-1}).
\end{equation}
\end{small}
GRU leverages a reset gate $r_t$ to control the input of the former state $h_{t-1}$:
\begin{small}
\begin{equation}\label{equ:reset}
r_t=\sigma(W_{r}E_{t}+U_{r}h_{t-1}).
\end{equation}
\end{small}
Then the activation of GRU is a linear interpolation between the previous activation $h_{t-1}$ and the candidate activation $\hat{h}_t$:
\begin{small}
\begin{equation}\label{equ:gru}
h_{t}=(1-z_{t})h_{t-1}+z_t \hat{h}_t,
\vspace{-0.5mm}
\end{equation}
\end{small}
where candidate activation function $\hat{h}_t$ is computed as:
\begin{small}
\begin{equation}\label{equ:gru1}
\hat{h}_t=\tanh[W E_{t}+U (r_t\cdot h_{t-1})].
\end{equation}
\end{small}
We use the final hidden state $h_{t}$ as the representation of the user's initial state $s^{ini}$ at the beginning of current recommendation session, i.e., $s^{ini} = h_{t}$.

\vspace{-2mm}
\subsubsection{\textbf{Encoder for Real-time State Generation Process}} 
\label{sec:local}
\begin{figure}[t]
	\centering
	\includegraphics[width=85mm]{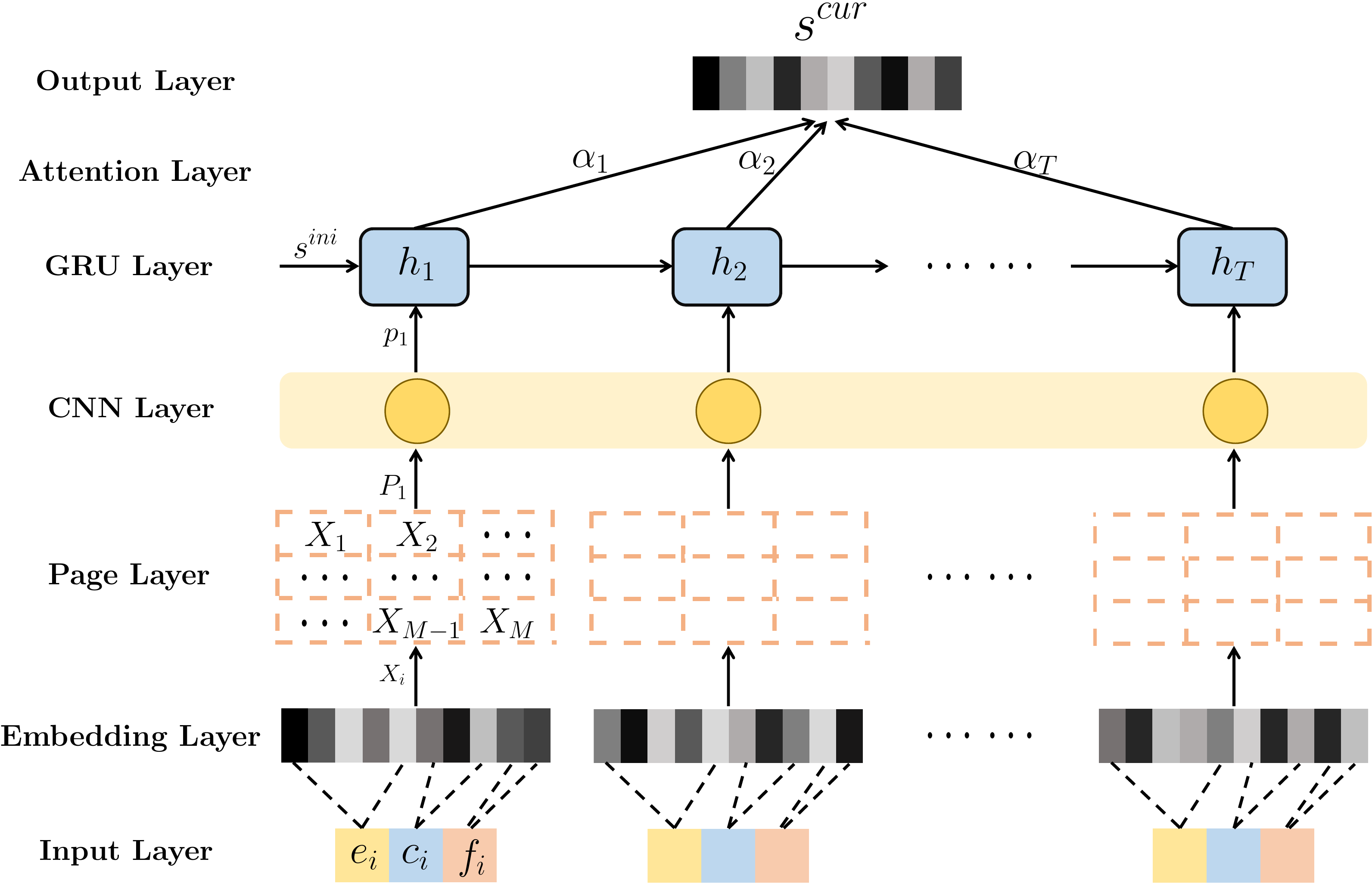}
	\caption{Encoder to generate real-time state $s^{cur}$.\label{fig:gru2}}
	\vspace{-6mm}
\end{figure}

Figure \ref{fig:gru2} illustrates the model to generate real-time preference in current session. In the page-wise recommender system, the inputs $\{x_1, \cdots, x_{M}\}$ for each recommendation page are the representations of the items in the page and user's corresponding feedback, where $M$ is the size of a recommendation page and $x_i$ is a tuple as:
\begin{small}
\begin{equation}\label{equ:tuple}
\vspace{-1.6mm}
x_i =(e_i,c_i,f_i),
\end{equation}
\end{small}
where $e_i$ is the aforementioned item representation. To assist the RA in capturing user's preference among different categories of items and generating complementary recommendations, we incorporate item's category $c_i$. The item's category $c_i$ is an one-hot indicator vector where $c_i(i) = 1$ if this item belongs to the $i^{th}$ category and other entities are zero. The one-hot indicator vector is extremely sparse and high-dimensional; hence we add a category-embedding layer transforming $c_i$ into a low-dimensional dense vector $C_i = \tanh(W_C c_i + b_C) \in \mathbb{R}^{|C|}$. 

In addition to information from items, $e_i$ and $c_i$, we also want to capture user's interests or feedback in current recommendation page. Thus, we introduce user's feedback vector $f_i$, which is an one-hot vector to indicate user's feedback for item $i$, i.e., skip/click/purchase.  Similarly, we transform $f_i$ into a dense vector $F_i = \tanh(W_F f_i + b_F) \in \mathbb{R}^{|F|}$ via the embedding layer. Finally, we get a low-dimensional dense vector $X_i \in \mathbb{R}^{|X|}$ ($|X| = |E| + |C| + |F|$) by concatenating $E_i$, $C_i$ and $F_i$ as:
\begin{small}
\begin{equation}\label{equ:concatenate}
\vspace{-1.6mm}
	\begin{aligned}
		X_i &=concat(E_i,C_i,F_i)\\		
		&=\tanh\big(concat(W_E e_i + b_E, W_C c_i + b_C, W_F f_i + b_F)\big).
	\end{aligned}
\end{equation}
\end{small}
\noindent Note that all item-embedding layers share the same parameters $W_E$ and $b_E$, which reduces the number of parameters and achieves better generalization. We apply the same constraints for category and feedback embedding layers.

Then, we reshape the transformed item representations $\{X_1, \cdots,$ $X_{M}\}$ as the original arrangement in the page. In other words, we arrange the item representations in one page $P_t$ as 2D grids similar to one image. For instance, if one recommendation page has $h$ rows and  $w$ columns ($M = h \times w$), we will get a $h \times w|X|$ matrix $P_t$. To learn item spatial display strategy in one page that leads to maximal reward, we introduce a Convolutional Neural Network (CNN).  CNN is a successful architecture in computer vision applications because of its capability to apply various learnable kernel filters on image to discover complex spatial correlations~\cite{krizhevsky2012imagenet}. Hence, we utilize 2D-CNN followed by fully connected layers to learn the optimal item display strategy as:
\begin{small}
\begin{equation}\label{equ:conv2d}
\vspace{-1mm}
p_t = conv2d(P_t),
\end{equation}
\end{small}
\noindent where $p_t$ is a low-dimensional dense vector representing the information from the items and user's feedback in page $P_t$ as well as the spatial patterns of the item display strategy of page $P_t$.

Next, we feed $\{p_1, \cdots, p_{T}\}$ into another RNN with Gated Recurrent Units (GRU) to capture user's real-time preference in the current session. The architecture of this GRU is similar to the one in Section \ref{sec:initial}, but we utilize the final hidden state $s^{ini}$ in Section \ref{sec:initial} as the initial state in current GRU. Furthermore, to capture the user's real-time preference in the current session, we employ attention mechanism~\cite{bahdanau2014neural}, which allows the RA to adaptively focus on and linearly combine different parts of the input sequence:
\begin{small}
\begin{equation}\label{equ:attention}
\vspace{-1.6mm}
s^{cur} = \sum_{t=1}^{T} \alpha_t h_t,
\end{equation}
\end{small}
\noindent where the weighted factors $\alpha_t$ determine which parts of the input sequence should be emphasized or ignored when making predictions. Here we leverage location-based attention mechanism~\cite{luong2015effective} where the weighted factors are computed from the target hidden state $h_t$ as follows:
\begin{small}
\begin{equation}\label{equ:weighted}
\vspace{-1.6mm}
\alpha_t= \frac{exp(W_{\alpha} h_t + b_{\alpha})}{\sum_{j}exp(W_{\alpha} h_j + b_{\alpha})}.
\end{equation}
\end{small}
This GRU with attention mechanism is able to dynamically select more important inputs, which is helpful to capture the user's real-time preference in the current session. Note that - 1) the length of this GRU is flexible according to that of the current recommendation session. After each user-agent interaction, i.e., user browse one page of generated recommendations and give feedback to RA, we can add one more GRU unit, and use this page of items, corresponding categories and feedback as the input of the new GRU unit; 2) in fact, the two RNNs in Section \ref{sec:initial} and Section \ref{sec:local} can be integrated into one RNN, we describe them separately to clearly illustrate their architecture, and clearly validate their effectiveness in Section \ref{sec:component}.

\vspace{-2mm}
\subsubsection{\textbf{Decoder for Action Generation Process}} 
\label{sec:action}
In this subsection, we will propose the action $a^{cur}$ generation process, which generates (recommends) a new page of items to users. In other words, given user's current preference $s^{cur}$, we aim to recommend a page of items and displays them properly to maximize the reward.  It is the inverse process of what the convolutional layer does. Hence, we use deconvolution neural network (DeCNN) to restore one page from the low-dimensional representation $s^{cur}$. It provides a sophisticated and automatic way to generate a page of recommendations with the corresponding display as:
\begin{small}
\begin{equation}\label{equ:decnn}
a^{cur} = deconv2d(s^{cur}).
\end{equation}
\end{small}
Note that - 1) the size of $a^{cur}$ and $P$ are different, since $a^{cur}$ only contains item-embedding $E_i$, while $P$ also contains item's category embedding $C_i$ and feedback-embedding $F_i$.  For instance, if one recommendation page has $h$ rows and $w$ columns ($M = h \times w$), $P$ is a $h \times w|X|$ matrix, while $a^{cur}$ is a $h \times w|E|$ matrix; and 2) the generated item embeddings in $a^{cur}$ may be not in the real item embedding set, thus we need to map them to valid item embeddings, which will be provided in later sections.
\vspace{-3mm}
\subsection{ The Architecture of Critic Framework}
\label{sec:traditional}
The Critic is designed to leverage an approximator to learn an action-value function $Q(s, a)$, which is a judgment of whether the action $a$ (or a recommendation page) generated by Actor matches the current state $s$. Note that we use ``$s$'' as $s^{cur}$ in the last subsection for simplicity.  Then, according $Q(s, a)$, the Actor updates its' parameters in a direction of improving performance to generate proper actions (or recommendations) in the following iterations.  

Thus we need to feed user's current state $s$ and action $a$ (or a recommendation page) into the critic. To generate user's current state $s$, the RA follows the same strategy from Equation (\ref{equ:update}) to Equation (\ref{equ:attention}), which uses embedding layers, 2D-CNN and GRU with attention mechanism to capture user's current preference. For action $a$, because $a^{cur}$ generated in Equation (\ref{equ:decnn}) is a 2D matrix similar to an image, we utilize the same strategy in Equation (\ref{equ:conv2d}), a 2D-CNN, to degrade $a^{cur}$ into a low-dimensional dense vector $a$ as:
\begin{small}
\begin{equation}\label{equ:conv2d1}
a = conv2d(a^{cur}).
\end{equation}
\end{small}
Then the RA concatenates current state $s$ and action $a$, and feeds them into a Q-value function $Q(s, a)$. In real recommender systems, the state and action spaces are enormous, thus estimating the action-value function $Q(s, a)$ for each state-action pair is infeasible. In addition, many state-action pairs may not appear in the real trace such that it is hard to update their values. Therefore, it is more flexible and practical to use an approximator function to estimate the action-value function. In practice, the action-value function is usually highly nonlinear. Thus we choose Deep neural networks as approximators. In this work, we refer to a neural network approximator as deep Q-value function (DQN). 
\vspace{-1.6mm}
\section{Training and Test Procedure}\label{sec:traini_test}
In this section, we discuss the training and test procedures.  We propose two polices, i.e., online-policy and off-policy, to train and test the proposed framework based on online environment and offline historical data, respectively.  Off-policy is necessary because the proposed framework should be pre-trained offline and be evaluated before launching them online to ensure the quality of the recommendations and mitigate possible negative impacts on user experience. After the offline stage, we can apply the framework online, and then the framework can continuously improve its strategies during the interactions with users.

\vspace{-1.6mm}
\subsection{The Training Procedure}
\label{sec:training}

\begin{figure}[t]
	\centering
	\includegraphics[width=81mm]{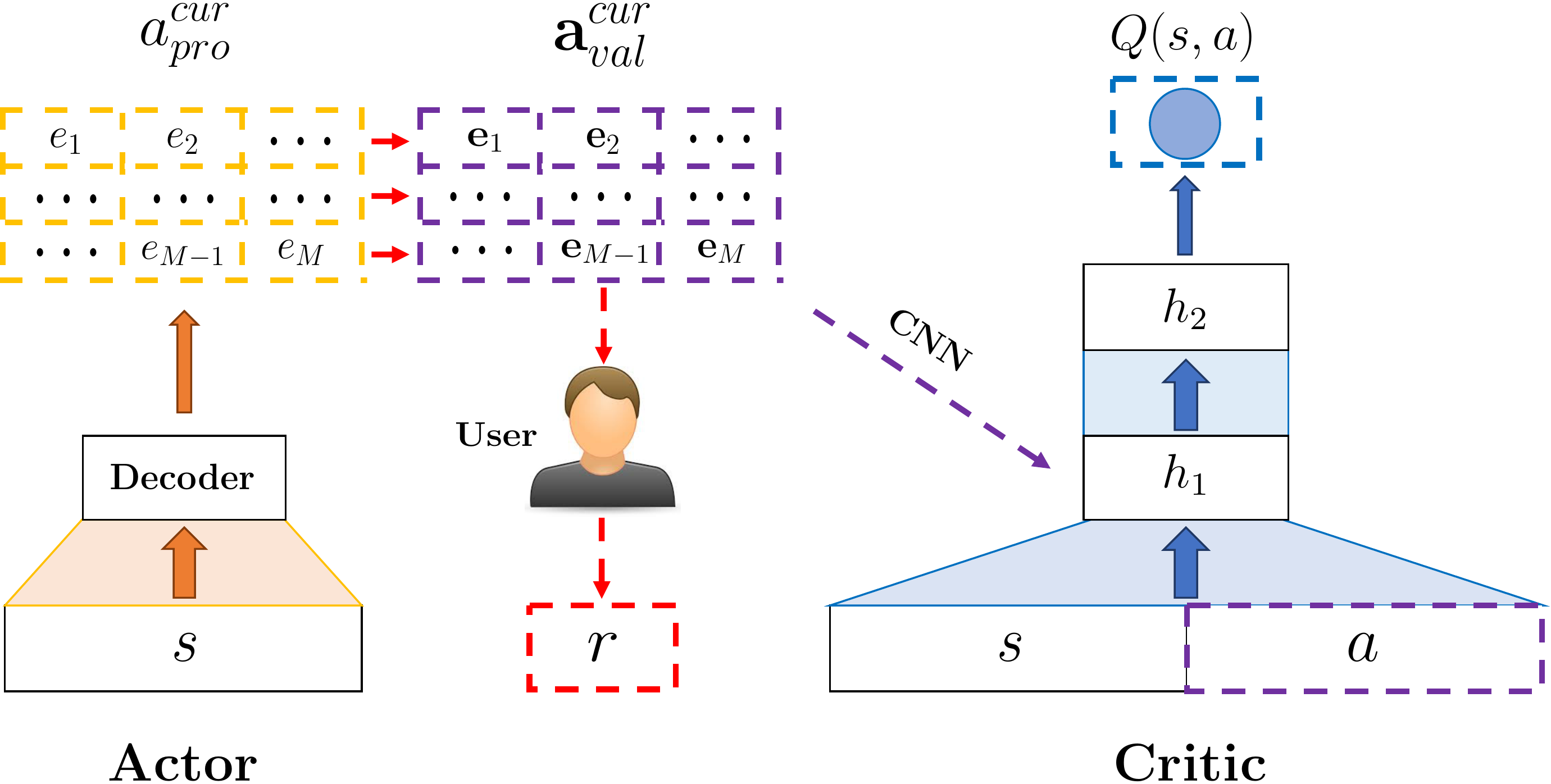}
	\caption{An illustration of the proposed framework.\label{fig:framework}}
	\vspace{-6mm}
\end{figure}

As aforementioned in Section \ref{sec:actor}, we map user's preference $s^{cur}$ to a new page of recommendations ($a^{cur}$). In a page of $M$ items, $a^{cur}$ contains item-embeddings of $M$ items, i.e., $\{e_1, \cdots, e_{M}\}$. However, $a^{cur}$ is a \textit{proto-action}, because the generated item embedding $e_i \in a^{cur}$ may be not in the existing item-embedding space $\mathcal{ I }$. Therefore, we need to map from proto-action $a_{pro}^{cur}$ to \textit{valid-action} $\mathbf{a}_{val}^{cur}$ where we have $\{\mathbf{e}_i \in \mathcal{ I } | \forall \mathbf{e}_i \in \mathbf{a}_{val}^{cur} \}$.  With this modification, an illustration of the proposed Actor-Critic recommending framework is demonstrated in Figure~\ref{fig:framework} where we omit Encoders for state generation part.

\vspace{-1.6mm}
\subsubsection{\textbf{Online Training Procedure}}
\label{sec:online}
When we train the proposed framework in online environment, RA can interact with users by sequentially choosing recommendation items over a sequence of time steps. Thus, in online environment, the RA is able to receive real-time feedback for the recommended items from users. In this setting, for each $e_i$ in $a_{pro}^{cur}$, we select the most similar $\mathbf{e}_i \in \mathcal{ I }$ as the \textit{valid item-embedding} in $\mathbf{a}_{val}^{cur}$. In this work, we select \textit{cosine similarity} as:
\vspace{-2mm}
\begin{small}
\begin{equation}\label{equ:cosine}
\mathbf{e}_i = \arg \max_{\mathbf{e} \in \mathcal{ I }} \frac{e_i^\top \cdot \mathbf{e}}{ \| e_i \| \| \mathbf{e} \|} = \arg \max_{\mathbf{e} \in \mathcal{ I }} \, e_i^\top \cdot \frac{ \mathbf{e}}{ \| \mathbf{e} \|}.
\end{equation}
\end{small}
To decrease the computational cost, we pre-compute $\frac{ \mathbf{e}}{ \| \mathbf{e} \|}$ and also adopt item recalling mechanism to reduce the number of relevant items~\footnote{ In general, user's preference in current session should be related to user's last clicked/purchased items before the current session(say $\mathcal{L}$). Thus for each item in $\mathcal{L}$, we collect a number of most similar items in terms of cosine similarity from the whole item space, and combine all collected items as the initial item-embedding space $\mathcal{ I }$ of current recommendation session. During the current session, when a user clicks or purchases an item, we will also add a number of its most similar items into the item-embedding space $\mathcal{ I }$.} . Note that Equation (\ref{equ:cosine}) is of the same complexity as the $\max_{a'}Q^{*}(s', a')$ in  Equation (\ref{equ:Q*sa}), $\mathcal{|A|}$, but each step of evaluation is a cosine similarity instead of a full value-function evaluation~\cite{dulac2015deep}.

\begin{algorithm}
	\small
	\caption{\label{alg:actor} Mapping Algorithm.}
	\raggedright
	{\bf Input}: User's browsing history, item-embedding space $\mathcal{I}$, the size of recommendation page $M$.\\
	{\bf Output}: Valid recommendation page  $\mathbf{a}_{val}^{cur}$.\\
	\begin{algorithmic} [1]
		\STATE Generate proto-action $a_{pro}^{cur}$ according \textbf{Eq.(\ref{equ:update})} to \textbf{Eq.(\ref{equ:attention})}
		\STATE \textbf{for} $m =1, M$ \textbf{do}
		\STATE \quad Select the most similar item as $\mathbf{e_m}$ according to \textbf{Eq.(\ref{equ:cosine})}
		\STATE \quad Add item $\mathbf{e_m}$ into $\mathbf{a}_{val}^{cur}$ (at the same location as $e_m$ in $a_{pro}^{cur}$)
		\STATE \quad Remove item $\mathbf{e_m}$ from $\mathcal{I}$
		\STATE \textbf{end for}
		\STATE \textbf{return} $\mathbf{a}_{val}^{cur}$
	\end{algorithmic}
\end{algorithm}
We present the mapping algorithm in Algorithm~\ref{alg:actor}. The Actor first generates proto-action $a_{pro}^{cur}$ (line 1). For each  $e_m$ in $a_{pro}^{cur}$, the RA selects the most similar item in terms of cosine similarity (line 3), and then adds this item into $\mathbf{a}_{val}^{cur}$ at the same position as $e_m$ in $a_{pro}^{cur}$ (line 4). Finally, the RA removes this item from the item-embedding space (line 5), which prevents recommending the same item repeatedly in one recommendation page. 

Then the RA recommends the new recommendation page $\mathbf{a}_{val}^{cur}$ to user, and receives the immediate feedback (reward) from user. The reward $r$ is the summation of rewards of all items in this page:
\begin{small}
\begin{equation}\label{equ:reward}
r = \sum_{m = 1}^{M} reward(\mathbf{e_m}).
\end{equation}
\end{small}
\vspace{-5mm}
\subsubsection{\textbf{Offline Training Procedure}}
\label{sec:offline}
When we use user's historical browsing data to train the proposed Actor-Critic framework, user's browsing history, the new recommendation page $\mathbf{a}_{val}^{cur}$ and user's corresponding feedback (reward) $r$ are given in the data. Thus, there is a gap between $a_{pro}^{cur}$ and $\mathbf{a}_{val}^{cur}$, i.e., no matter what proto-action $a_{pro}^{cur}$ outputted by the Actor, the valid-action $\mathbf{a}_{val}^{cur}$ is fixed. This will disconnect the Actor and the Critic.

From existing work~\cite{lillicrap2015continuous, dulac2015deep} and Section \ref{sec:online}, we learn that $a_{pro}^{cur}$ and $\mathbf{a}_{val}^{cur}$ should be similar, which is the prerequisite to connect the Actor and the Critic for training. Thus, we choose to minimize the difference between $a_{pro}^{cur}$ and $\mathbf{a}_{val}^{cur}$:
\begin{small}
\begin{equation}\label{equ:loss_function}
\vspace{-1.6mm}
\min_{\theta^{\pi}} \sum_{b=1}^B \big( \| a_{pro}^{cur} - \mathbf{a}_{val}^{cur} \|_F^2 \big),
\end{equation}
\end{small}
where $B$ is the batch size of samples in each iteration of SGD. Equation(\ref{equ:loss_function}) updates Actor's parameters in the direction of pushing $a_{pro}^{cur}$ and $\mathbf{a}_{val}^{cur}$ to be similar. In each iteration, given user's browsing history, the new recommendation page $\mathbf{a}_{val}^{cur}$, the RA generates proto-action $a_{pro}^{cur}$ and then minimizes the difference between $a_{pro}^{cur}$ and $\mathbf{a}_{val}^{cur}$, which can connect the Actor and the Critic. Next, we can follow conventional methods to update the parameters of Actor and Critic. The reward $r$ is the summation of rewards of all items in page $\mathbf{a}_{val}^{cur}$.
\vspace{-1.6mm}
\subsubsection{\textbf{Training Algorithm}}
\label{sec:trainingA}
In this work, we utilize DDPG \cite{lillicrap2015continuous} algorithm to train the parameters of the proposed Actor-Critic framework. The Critic can be trained by minimizing a sequence of loss functions $L(\theta^\mu)$ as:
\begin{small}
\begin{equation}\label{equ:L}
\vspace{-1mm}
L(\theta^\mu)=\mathbb{E}_{s, a,r,s'}\big[\big(r+\gamma Q_{\theta^{\mu'}}(s',f_{\theta^{\pi'}}(s'))-Q_{\theta^\mu}(s, a)\big)^{2}\big],
\end{equation}
\end{small}
where $\theta^{\mu}$ represents all parameters in Critic. The critic is trained from samples stored in a replay buffer~\cite{mnih2015human}. Actions stored in the replay buffer are generated by valid-action $\mathbf{a}_{val}^{cur}$, i.e., $a = conv2d(\mathbf{a}_{val}^{cur})$. This allows the learning algorithm to leverage the information of which action was actually executed to train the critic~\cite{dulac2015deep}. 

The first term $y = r+\gamma Q_{\theta^{\mu'}}(s',f_{\theta^{\pi'}}(s'))$ in Equation (\ref{equ:L}) is the target for the current iteration. The parameters from the previous iteration $\theta^{\mu'}$ are fixed when optimizing the loss function $L(\theta^\mu)$. In practice, it is often computationally efficient to optimize the loss function by stochastic gradient descent, rather than computing the full expectations in the above gradient. The derivatives of loss function $L(\theta^\mu)$ with respective to parameters $\theta^\mu$ are presented as follows:
\begin{small}
\begin{equation}\label{equ:differentiating}
\vspace{-2.1mm}
\begin{aligned}
\nabla_{\theta^\mu}L(\theta^\mu) & =\mathbb{E}_{s, a,r,s'}\big[(r+\gamma Q_{\theta^{\mu'}}(s',f_{\theta^{\pi'}}(s'))\\
&-Q_{\theta^\mu}(s, a))\nabla_{\theta^\mu}Q_{\theta^\mu}(s, a)\big].
\end{aligned}
\end{equation}
\end{small}
We update the Actor using the policy gradient:
\begin{small}
\begin{equation}\label{equ:policy}
\vspace{-1mm}
\nabla_{\theta^{\pi}}f_{\theta^\pi}  \approx  \mathbb{E}_{s} \big[\nabla_{\hat{a}}Q_{\theta^\mu}(s, \hat{a}) \,   \nabla_{\theta^{\pi}}f_{\theta^\pi}(s)\big],
\end{equation}
\end{small}
where $\hat{a} = f_{\theta^\pi}(s)$, i.e., $\hat{a}$ is generated by proto-action $a_{pro}^{cur}$ ($\hat{a} = conv2d(a_{pro}^{cur})$). Note that proto-action $a_{pro}^{cur}$ is the actual action outputted by Actor. This guarantees that policy gradient is taken at the actual output of policy $f_{\theta^\pi}$~\cite{dulac2015deep}.
\begin{algorithm}
	\small
	\caption{\label{alg:training} Parameters Online Training for DeepPage with DDPG.}
	\raggedright
	\begin{algorithmic} [1]
		\STATE Initialize actor network $f_{\theta^\pi}$ and critic network $Q_{\theta^{\mu}}$ with random weights	
		\STATE Initialize target network $f_{\theta^{\pi'}}$ and $Q_{\theta^{\mu'}}$ with weights $\theta^{\pi'}\leftarrow\theta^{\pi}, \theta^{\mu'}\leftarrow\theta^{\mu}$
		\STATE Initialize the capacity of replay buffer $\mathcal{D}$
		\FOR{$session =1, G$}
		\STATE  Receive initial observation state $s_1$
		\FOR{$t=1, T$}
		\STATE \textbf{Stage 1: Transition Generating Stage}
		\STATE  Select an action $a_t$ according to \textbf{Alg.\ref{alg:actor}} (policy $f_{\theta^{\pi}}$)
		\STATE  Execute action $a_t$ and observe the reward $r_t$ according to \textbf{Eq. (\ref{equ:reward})} and new state $s_{t+1}$ according to \textbf{Section \ref{sec:local}}
		\STATE  Store transition $(s_t, a_t,  r_t,s_{t+1})$ in $\mathcal{D}$
		\STATE \textbf{Stage 2: Parameter Updating Stage}
		\STATE  Sample minibatch of $\mathcal{N}$ transitions $(s, a, r, s')$ from $\mathcal{D}$
		\STATE  Set $y= r+\gamma Q_{\theta^{\mu'}}(s',f_{\theta^{\pi'}}(s'))  $
		\STATE  Update Critic by minimizing $\frac{1}{\mathcal{N}}\sum_{n}\big(y-Q_{\theta^\mu}(s, a)\big)^{2}$ according to \textbf{Eq. (\ref{equ:differentiating}) }
		\STATE  Update Actor using the sampled policy gradient according to \textbf{Eq. (\ref{equ:policy}) }
		\STATE Update the target networks: $$\theta^{\mu'}\leftarrow\tau\theta^{\mu}+(1-\tau)\theta^{\mu'}$$
		$$	\theta^{\pi'}\leftarrow\tau\theta^{\pi}+(1-\tau)\theta^{\pi'}	$$
		\ENDFOR 
		\ENDFOR 
	\end{algorithmic}
\end{algorithm}

The online training algorithm for the proposed framework DeepPage is presented in Algorithm \ref{alg:training}. In each iteration, there are two stages, i.e., 1) transition generating stage (lines 7-10), and 2) parameter updating stage (lines 11-16). For transition generating stage (line 7): given the current state $s_t$, the RA first recommends a page of items $a_t$ according to Algorithm~\ref{alg:actor} (line 8); then the RA observes the reward $r_t$ and updates the state to $s_{t+1}$ (lines 9); and finally the RA stores transitions $(s_t,a_t,r_t,s_{t+1})$ into the replay buffer $\mathcal{D}$ (line 10). For parameter updating stage (line 11): the RA samples mini-batch of transitions $(s, a, r, s')$ from $\mathcal{D}$ (line 12), and then updates parameters of Actor and Critic (lines 13-16) following a standard DDPG procedure~\cite{lillicrap2015continuous}.

The offline training procedure is similar with Algorithm~\ref{alg:training}. The two differences are: 1) in line 8, offline training follows off-policy $b(s_t)$, and 2) before line 13, offline training first minimizes the difference between $a_{pro}^{cur}$ and $\mathbf{a}_{val}^{cur}$ according to Equation (\ref{equ:loss_function}). 
\vspace{-2mm}
\subsection{The Test Procedure}
\label{sec:test}
After the training stage, the proposed framework learns parameters $\Theta^{\pi}$ and $\Theta^{\mu}$. Now we formally present the test procedure of the proposed framework DeepPage. We design two test methods, i.e., 1) Online test: to test DeepPage in online environment where RA interacts with users and receive real-time feedback for the recommended items from users, and 2) Offline test: to test DeepPage based on user's historical browsing data.

\vspace{-2mm}
\subsubsection{\textbf{Online Test}}
\label{sec:on_test}
The online test algorithm in one recommendation session is presented in Algorithm~\ref{alg:test_on}. The online test procedure is similar with the transition generating stage in Algorithm~\ref{alg:training}.  In each iteration of the recommendation session, given the current state $s_t$, the RA recommends a page of recommendations $a_t$ to user following policy $f_{\Theta^{\pi}}$ (line 4). Then the RA observes the reward $r_t$ from user (line 5) and updates the state to $s_{t+1}$ (line 6). 
\vspace{-2.1mm}
\begin{algorithm}
	\small
	\caption{\label{alg:test_on}Online Test for DeepPage.}
	\raggedright
	\begin{algorithmic} [1]
		\STATE Initialize Actor with the trained parameters $\Theta^{\pi}$
		\STATE  Receive initial observation state $s_1$
		\FOR{$t=1, T$}
		\STATE  Execute an action $a_t$ according to \textbf{Alg.\ref{alg:actor}} (policy $f_{\Theta^{\pi}}$)   
		\STATE  Observe the reward $r_t$ from user according to \textbf{Eq. (\ref{equ:reward})} 
		\STATE  Observe new state $s_{t+1}$ according to \textbf{Section \ref{sec:local}}
		\ENDFOR 
	\end{algorithmic}
\end{algorithm}

\vspace{-1mm}
\subsubsection{\textbf{Offline Test}}
\label{sec:off_test}
The intuition of our offline test method is that, for a given recommendation session (offline data), the RA reranks the items in this session. If the proposed framework works well, the clicked/purchased items in this session will be ranked at the top of the new list. The reason why RA only reranks items in this session rather than items in the whole item space is that for the offline dataset, we only have the ground truth rewards of the existing items in this session. The offline test algorithm in one recommendation session is presented in Algorithm \ref{alg:test_off}. In each iteration of an offline test recommendation session, given the state $s_t$ (line 2), the RA recommends an page of recommendations $a_t$ following policy $f_{\Theta^{\pi}}$ (lines 4). For each item $e_i$ in $a_t$, we add it into new recommendation list $\mathcal{L}$ (line 6), and record $e_i$'s reward $r_i$ from user's historical browsing data (line 7). Then we can compute the overall reward $r_t$ of $a_t$ (line 9) and update the state to $s_{t+1}$ (line 10). Finally, we remove all items $e_i$ in $a_t$ from the item set $\mathcal{I}$ of the current session (line 11).
\vspace{-3mm}
\begin{algorithm}
	\small
	\caption{\label{alg:test_off}Offline Test of DeepPage Framework.}
	\raggedright
	{\bf Input}: Item set $\mathcal{I} = \{ \mathbf{e}_1, \cdots, \mathbf{e}_N\}$ and corresponding reward set $\mathcal{R} = \{r_1, \cdots, r_N\}$  of a session.\\
	{\bf Output}:Recommendation list $\mathcal{L}$  with new order\\
	\begin{algorithmic} [1]
		\STATE Initialize Actor with well-trained parameters $\Theta^{\pi}$
		\STATE  Receive initial observation state $s_1$
		\WHILE{$|\mathcal{I}| > 0$}
		\STATE  Select an action $a_t$ according to \textbf{Alg.\ref{alg:actor}} (policy $f_{\Theta^{\pi}}$)  
		\FOR{$e_i \in a_t$}
		\STATE  Add $e_i$ into the end of $\mathcal{L}$
		\STATE  Record reward $r_i$ from user's historical browsing data
		\ENDFOR
		\STATE  Compute the overall reward $r_t$ of $a_t$ according to \textbf{Eq. (\ref{equ:reward})} 
		\STATE  Execute action $a_t$ and observe new state $s_{t+1}$ according to \textbf{Section \ref{sec:local}}
		\STATE  Remove all $e_i \in a_t$ from $\mathcal{I}$ 
		\ENDWHILE
	\end{algorithmic}
\end{algorithm}
\vspace{-3mm}



\vspace{-2.1mm}
\section{Experiments}\label{sec:experiments}
In this section, we conduct extensive experiments with a dataset from a real e-commerce company to evaluate the effectiveness of the proposed framework. We mainly focus on two questions: (1) how the proposed framework performs compared to representative baselines; and (2) how the components in Actor and Critic contribute to the performance. 
\vspace{-1.6mm}
\subsection{Experimental Settings}
\label{sec:experimental_settings}
We evaluate our method on a dataset of September, 2017 from a real e-commerce company. We randomly collect 1,000,000 recommendation sessions (9,136,976 items) in temporal order, and use the first 70\% sessions as the training/validation set and the later 30\% sessions as the test set. For a new session, the initial state is collected from the previous sessions of the user. In this work, we leverage $N = 10$ previously clicked/purchased items to generate the initial state. Each time the RA recommends a page of $M = 10$ items (5 rows and 2 columns) to users~\footnote{This is based on offline historical data collected from mobile App, i.e., to fit the screen size of mobile phones, one page has only 5 rows and 2 columns.}. The reward $r$ of one skipped/clicked/purchased item is empirically set as 0, 1, and 5, respectively. The dimensions of item-embedding/category-embedding/ feedback-embedding are $|E| = 50$, $|C| = 35$, and $|F| = 15$. We set the discounted factor $\gamma = 0.95$, and the rate for soft updates of target networks $\tau = 0.01$. For offline test, we select \textbf{Precision@20}, \textbf{Recall@20}, \textbf{F1-score@20}~\cite{gunawardana2009survey}, \textbf{NDCG@20}~\cite{jarvelin2002cumulated} and \textbf{MAP}~\cite{turpin2006user},  as the metrics.  For online test, we leverage the summation of all rewards in one recommendation session as the metric. 

\begin{figure}[t]
	\centering
	\includegraphics[width=81mm]{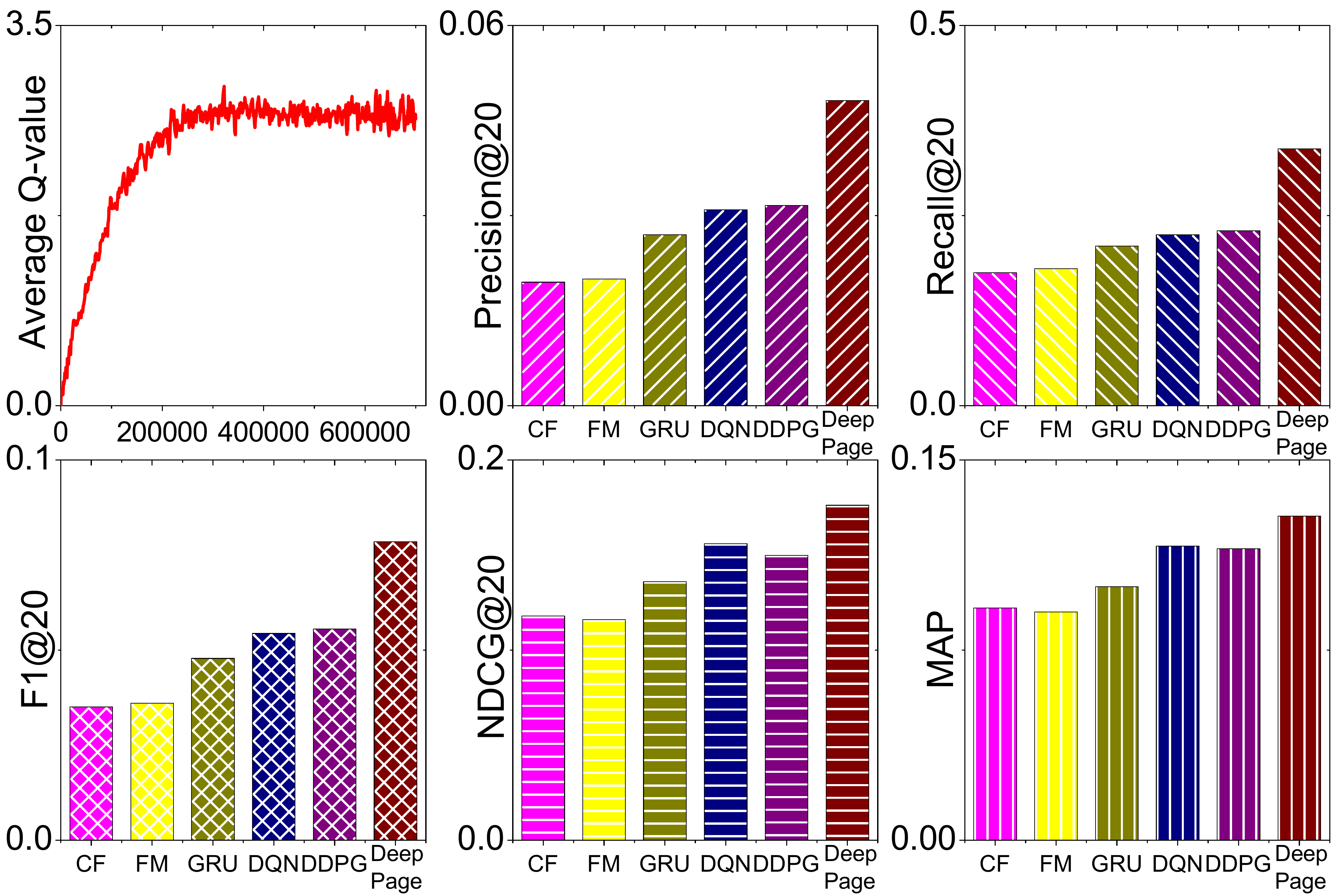}
	\vspace{-2mm}
	\caption{Overall performance comparison in offline test.}
	\label{fig:overall}
	\vspace{-6mm}
\end{figure}
\vspace{-2.1mm}
\subsection{Performance Comparison for Offline Test}
\label{sec:ev_overall}
To answer the first question, we compare the proposed framework with the following representative baseline methods -- \textbf{CF}~\cite{breese1998empirical}: collaborative filtering; \textbf{FM}~\cite{rendle2010factorization}: Factorization Machines; \textbf{GRU}~\cite{hidasi2015session} : it utilizes the Gated Recurrent Units (GRU) to predict what user will click/purchase next based on the browsing histories (to make a fair comparison, it also keeps previous $N = 10$ clicked/purchased items as initial states);  \textbf{DQN}~\cite{mnih2013playing}: we use a Deep Q-network with five fully-connected layers in this baseline. The input is the concatenation of embeddings of users' historical clicked/purchased items (state) and a page of recommendations (action), and train this baseline by Eq.(\ref{equ:Q*sa}); and \textbf{DDPG}~\cite{dulac2015deep}: in this baseline, we use conventional Deep Deterministic Policy Gradient with five fully connected layers in both Actor and Critic. 

We leverage offline training strategy to train DDPG and DeepPage as mentioned in Section \ref{sec:online}. The results are shown in Figure~\ref{fig:overall}. Figure \ref{fig:overall} (a) illustrates the training process of DeepPage. We can observe that the framework approaches convergence when the model is trained by 500,000 offline sessions. We note that  CF and FM perform worse than other baselines. These two baselines ignore the temporal sequence of the users' browsing history, while GRU can capture the temporal sequence, and DQN, DDPG and DeepPage are able to continuously update their strategies during the interactions. DQN and DDPG outperform GRU.  We design GRU to maximize the immediate reward for recommendations, while DQN and DDPG are designed to achieve the trade-off between short-term and long-term rewards. This result suggests that introducing reinforcement learning can improve the performance of recommendations. Finally DeepPage performs better than conventional DDPG.  Compared to DDPG, DeepPage jointly optimizes a page of items and uses GRU to learn user's real-time preference. 

\vspace{-2.1mm}
\subsection{Performance Comparison for Online Test}
Following~\cite{zhao2018recommendations}, we build a simulated online environment (adapted to our case) for online test. We compare DeepPage with GRU, DQN and DDPG. Here we utilize online training strategy to train DDPG and DeepPage (both Actor-Critic framework) as mentioned in Section \ref{sec:offline}. Baselines are also applicable to be trained via the rewards generated by simulated online environment. Note that we use data different from the training set to build the simulated online environment to avoid over-fitting. 

\begin{figure}[t]
	\centering
	\includegraphics[width=81mm]{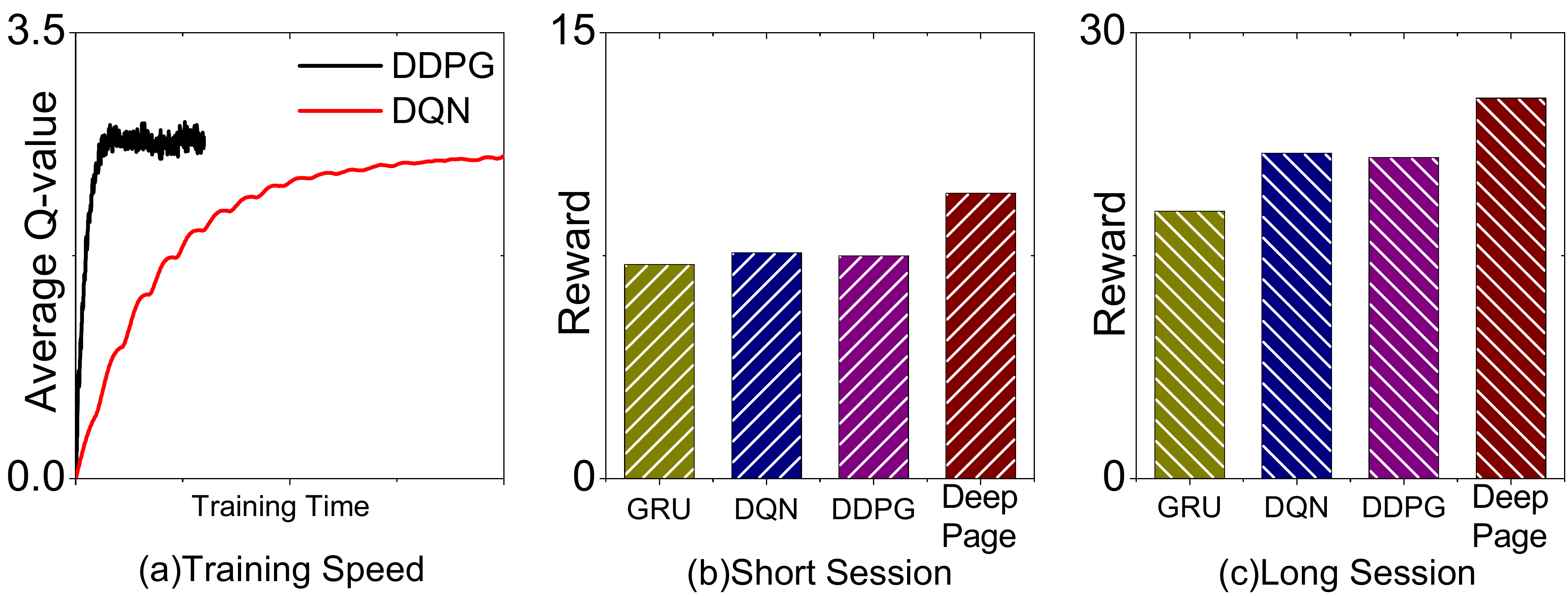}
	\vspace{-2mm}
	\caption{Overall performance comparison in online test.}
	\label{fig:online}
	\vspace{-3mm}
\end{figure}

As the test stage is based on the simulator, we can artificially control the length of recommendation sessions to study the performance in short and long sessions. We define short sessions with 10 recommendation pages, while long sessions with 50 recommendation pages. The results are shown in Figure \ref{fig:online}. We note DDPG performs similar to DQN, but the training speed of DDPG is much faster than DQN, as shown in Figure \ref{fig:online} (a).  This result indicates that Actor-Critic framework is suitable for practical recommender systems with enormous action space. In short recommendation sessions, GRU, DQN and DDPG achieve comparable performance. In other words, GRU model and reinforcement learning models like DQN and DDPG can recommend proper items matching users' short-term interests. In long recommendation sessions, DQN and DDPG outperform GRU significantly.  GRU is designed to maximize the immediate reward for recommendations, while DQN and DDPG are designed to achieve the trade-off between short-term and long-term rewards. This result suggests that introducing reinforcement learning can improve the long-term performance of recommendations.  Finally DeepPage performs better than conventional DQN and DDPG. DeepPage can learn user's real-time preference and optimize a page of items. We detail the effect of model components of DeepPage in the following subsection. 

\vspace{-3mm}
\subsection{Effectiveness of Components}
\label{sec:component}
To validate the effectiveness of each component, we systematically eliminate the corresponding model components by defining following variants of DeepPage -- (1) \textbf{DeepPage-1}: this variant is to evaluate the performance of the embedding layers. We remove the embedding layers for items, categories and feedback; (2) \textbf{DeepPage-2}: in this variant, we evaluate the contribution of category and feedback information, hence, this variant does not contain one-hot indicator vectors of category and feedback; (3) \textbf{DeepPage-3}: This variant is to evaluate the effectiveness of GRU to generate initial state, so we eliminate the GRU in Figure \ref{fig:gru1}; (4) \textbf{DeepPage-4}: In this variant, we evaluate the contribution of CNNs as shown in Figure \ref{fig:gru2}, thus we remove CNNs and directly feed the outputs of embedding layers (concatenate embeddings of all items as one vector) into GRU units; (5) \textbf{DeepPage-5}: This variant is to evaluate the effectiveness of attention mechanism in Figure \ref{fig:gru2}, therefore, we eliminate attention layer and use the hidden state of last GRU unit as the input of DeCNN; (6) \textbf{DeepPage-6}: In this variant, we evaluate the GRU to generate real-time state, thereby, we remove this GRU in Figure \ref{fig:gru2} and concatenate outputs of all CNNs as a vector, and feed it into DeCNN; and (7) \textbf{DeepPage-7}: This variant is to evaluate the performance of DeCNN to generate a new page of items, hence, we replace it by fully-connected layers, which output a concatenated vector of $M$ item-embeddings.

The offline results are shown in Table \ref{table:component}. Note that we omit similar online observations because of the space limitation. DeepPage-1 and DeepPage-2 validate that incorporating category/feedback information and the embedding layers can boost the performance. DeepPage-3 and DeepPage-6 perform worse, which suggests that setting user's initial preference at the beginning of a new recommendation session, and capturing user's real-time preference in current session is helpful for accurate recommendations.  DeepPage-5 proves that incorporating attention mechanism can better capture user's real-time preference than only GRU. DeepPage outperforms DeepPage-4 and DeepPage-7, which indicates that item display strategy can influence the decision making process of users. In a nutshell, DeepPage outperforms all its variants, which demonstrates the effectiveness of each component for recommendations.

\begin{tiny}
	\begin{table}[t]
		\centering
		\small
		\caption{Performance comparison of different components.}
		\begin{tabular}{p{1.56cm}<{\centering}||p{0.966cm}<{\centering}|p{0.966cm}<{\centering}|p{0.966cm}<{\centering}|p{0.966cm}<{\centering}|p{0.966cm}<{\centering}}
			\hline
			{\large Methods}& Precision& Recall& F1score& NDCG& MAP   \\
			\textbf{}& @20 & @20 & @20 & @20 &    \\
			\hline\hline
			DeepPage-1  & 0.0479 & 0.3351 & 0.0779 & 0.1753 & 0.1276  \\
			DeepPage-2 & 0.0475 & 0.3308 & 0.0772 & 0.1737 & 0.1265  \\
			DeepPage-3 & 0.0351 & 0.2627 & 0.0578 & 0.1393 & 0.1071   \\
			DeepPage-4 & 0.0452 & 0.3136 & 0.0729 & 0.1679 & 0.1216   \\
			DeepPage-5 & 0.0476 & 0.3342 & 0.0775 & 0.1716 & 0.1243   \\
			DeepPage-6 & 0.0318 & 0.2433 & 0.0528 & 0.1316 & 0.1039   \\
			DeepPage-7 & 0.0459 & 0.3179 & 0.0736 & 0.1698 & 0.1233   \\
			\textbf{DeepPage}     & \textbf{0.0491} & \textbf{0.3576} & \textbf{0.0805} & \textbf{0.1872} & \textbf{0.1378} \\ \hline
		\end{tabular}
		\label{table:component}
		\vspace{-3mm}
	\end{table}
\end{tiny}

\vspace{-3mm}
\section{Related Work}
\label{sec:related_work}
In this section, we briefly review research related to our study, i.e., reinforcement learning for recommendations. The MDP-Based CF model can be viewed as approximating a partial observable MDP \cite{shani2005mdp}. Mahmood et al.\cite{mahmood2009improving} adopted the RL technique to observe the responses of users in a conversational recommender. Taghipour et al.\cite{taghipour2007usage,taghipour2008hybrid} modeled web page recommendation as a Q-Learning problem and learned to make recommendations from web usage data. Sunehag et al.\cite{sunehag2015deep} addressed sequential decision problems with high-dimensional slate-action spaces. Zheng et al.\cite{zheng2018drn} proposed a RL framework to do online news recommendation. Cai et al.\cite{cai2018reinforcement1,cai2018reinforcement2} employed a RL mechanism for impression allocation problem. Feng et al.~\cite{feng2018learning} optimized ranking strategies collaboratively for multi-scenario recommendations. Choi et al.~\cite{choi2018reinforcement} used RL and biclustering for cold-start problem. Chen et al.~\cite{chen2018stabilizing} proposed strategies to improve the reward estimation in dynamic recommendation environments. Zhao et al.~\cite{zhao2018recommendations} captured both positive and negative feedback into RL based recommendations. Zhao et al.~\cite{zhao2017deep} optimized a set of recommendations with proper order.

\vspace{-2.1mm}
\section{Conclusion}
\label{sec:conclusion}

In this paper, we propose a novel page-wise recommendation framework, DeepPage, which leverages Deep Reinforcement Learning to automatically learn the optimal recommendation strategies and optimizes a page of items simultaneously. We validate the effectiveness of our framework with extensive experiments based on data from a real e-commerce company. Further research directions include reducing the temporal complexity of mapping from proto-action to valid-action and handling multiple tasks such as search, bidding, advertisement and recommendation collaboratively in one reinforcement learning framework.

\vspace{-2.1mm}
\section*{Acknowledgements}
This material is based upon work supported by the National Science Foundation (NSF) under grant number IIS-1714741 and IIS-1715940, and a grant from Criteo Faculty Research Award. 
	
\bibliographystyle{ACM-Reference-Format}
\bibliography{9reference} 

\end{document}